# The gravitational waves are fictitious entities

A. LOINGER

Dipartimento di Fisica, Università di Milano

Via Celoria, 16, 20133 Milano, Italy

ABSTRACT. – The gravitational waves are non-physical sinuosities generated, in the last analysis, by undulating reference frames.

KEY WORDS: Gravitational radiation; Characteristics and wave fronts; Developable hypersurfaces.

PACS. 04.30 – Gravitational waves and radiation: theory.

**1.** – We show that the gravitational waves of general relativity are only analytical sinuosities generated by purely formal approaches.

First of all, the *emission* mechanism of the gravitational waves is a real mystery. It is commonly asserted that any accelerated mass point must give out gravitational waves (cf. e.g. Bergmann, 1960, p.187), but in general relativity acceleration does not have an intrinsic, absolute meaning (and the metric $g_{ik}$ in an accelerated frame gives a curvature tensor equal to zero). Weyl wrote (1988, p.248): "*Jede Änderung der Materieverteilung bringt eine Gravitationswirkung hervor, die sich im Raum mit Lichtgeschwindigkeit ausbreitet.* Schwingende Massen erzeugen Gravitationswellen." Now, this conclusion was derived by a linear (or weak-field) approximation of Einstein's equations and by using harmonic co-ordinates; in this approximation we put $g_{ik} \approx \eta_{ik} + \varepsilon_{ik}$, i.e. we consider the world as almost pseudo-Euclidean and characterized essentially by a Minkowskian fundamental tensor $\eta_{ik}$, while the $\varepsilon_{ik}$'s are the components of a small perturbing tensor field. In *this* schematization the accelerated physical systems are privileged. However, "it is well known that the solutions of non-linear equations cannot be fairly approximated by solutions of the linear equations over larger parts of space-time." (cf. e.g. Möller, 1972, p.465). (See also Einstein and Rosen, 1937 and the perturbation treatments by Blanchet and Damour, 1986, Blanchet, 1987 and Damour and Schmidt, 1990).

Thus, from the rigorous point of view of the non-linear theory *there is no mechanism for the emission of gravitational waves*. Further, as it will be clear in the sequel, the more significant wave fronts of Einstein's equations are *developable* hypersurfaces, on which a *constant* metric can be obviously chosen. Last but not least, since the gravitational pseudo energy cannot be localized, the *absorption* mechanism of a hypothetic gravitational wave by an experimental receptor is not very transparent.

Remark also that we are not authorized to discard *a priori* the ingoing waves and suitable mixtures of ingoing and outgoing waves (cf. e.g. Einstein and Rosen, 1937).

**2.** − Levi-Civita (1930) (see also Straumann, 1984, p.163) proved rigorously − and avoiding the use of special co-ordinates, as e.g. the harmonic co-ordinates (cf. e.g. Fock, 1964, p.192 and p.342) − that from the exact non-linear Einstein's field equations the following differential equation for the functions $z(x)$ of the characteristic hypersurfaces $z(x) = 0$, $[x: = (x^0, x^1, x^2, x^3)]$, can be deduced ($i, k = 0, 1, 2, 3$):

$$(2.1) \qquad g^{ik}(x) \frac{\partial z(x)}{\partial x^i} \frac{\partial z(x)}{\partial x^k} = 0 .$$

The physical meaning of eq. (2.1), which coincides formally with the characteristic equation of d'Alembert equation, is that $z(x) = 0$ gives in general the law of motion of a *wave front*. The differential equations of the bicharacteristics, i.e. of the characteristic lines of (2.1) − which can be re-written in a Hamilton-Jacobi form

(2.1')  $\quad H := \tfrac{1}{2} g^{ik} p_i p_k = 0$ ,  with  $p_i := \partial z / \partial x^i$  − ,

are the following Hamiltonian equations (with the specification that $H = 0$ for all the interesting solutions):

(2.2)  $\quad \dfrac{dp_i}{d\sigma} = -\dfrac{\partial H}{\partial x^i}$ ,  $\quad \dfrac{dx^i}{d\sigma} = \dfrac{\partial H}{\partial p_i}$ ,

where $\sigma$ is an auxiliary parameter. These characteristic lines − the *rays* of the wave front $z(x) = 0$ − coincide with the null geodesic $ds = 0$ , with $ds^2 = g_{ik} \, dx^i \, dx^k$ .

Now, we have the well-known theorem of Fermi (1922) (see also Levi-Civita, 1925, p.190): on any line segment of a Riemannian manifold it is possible to choose a co-ordinate system which is locally geodesic on this segment. Intuitively: the envelope of the tangent hyperplanes to the line points is a developable hypersurface; etc.. (Fermi's theorem has been extended by Eisenhart to the non-Riemannian manifolds with a symmetric connection, cf. e.g. Eisenhart, 1927, p.64).

We apply this theorem to any solution of eqs. (2.2), and thus we can impress a *constant* metric on a given ray. This result can be immediately extended to a whole *characteristic hypersurface* whose rays are *parallel* (according to space-time geometry). And it holds also for all the waves whose characteristic is a **developable** hypersurface, as it is clear. Example: cylindrical waves in a three-dimensional Euclidean *Bildraum*, as in the exact solution by Einstein and Rosen (1937). Finally, since at a sufficiently great distance from the origin a large portion of *any* gravitational wave can be considered as *plane*, the above conclusion has a rather general extent. Levi-Civita (1930) does *not* even mention the gravitational waves and refers eqs. (2.1) – (2.1'), (2.2) to the *electromagnetic* field. See also Whittaker (1927) and Fock (1964, p.433).

**3.** – In Fock's treatise (1964, p.398) we find a remark which is particularly interesting for our thesis. With reference to the well-known Einstein-Infeld-Hoffmann method for deriving from Einstein's field equations – in an explicit form, with successive approximations – the equations of motion of point-like singularities of the metric tensor, Fock writes: "In those cases when approximations higher than the second are discussed another question arises ..... It becomes unclear whether the formal solution obtained has physical sense. This doubt is due to the fact that the authors of these papers do not impose the condition of outward radiation [the *Ausstrahlungsbedingung* of the German authors] and, in fact, make the contrary assertion that there exist certain (unknown) co-ordinate transformations which reduce the exact equations of motion to Newtonian form, corresponding to a strictly conservative non-radiating system. Both parts of the latter assertion, namely the possibility of reducing more general equations of motion to Newtonian form and the possibility of reducing a non-conservative (radiating) system to a conservative one, seem doubtful."

This criticism does not seem appropriate. Indeed, the possibility that at each step of the EIH method the contributions of the gravitational radiation can be eliminated through a suitable co-ordinate transformation, is immediately comprehensible if one observes that, by virtue of our previous considerations, the gravitational waves are only artificial sinuosities.

**4.** – Finally, we ask ourselves: if the gravitational waves are fictitious entities, which is the real physical meaning of eqs. (2.1) – (2.1'), (2.2) ?
We have pointed out that Levi-Civita (1930) refers these equations to the *electromagnetic* wave fronts and rays. At first sight this seems a heresy, still more because eqs. (2.1) − (2.1'), (2.2) follow from Einstein's field equations even if the mass tensor is equal to zero. But it is easily seen that Levi-Civita's interpretation is quite sound. In *special* relativity the differential equations of the characteristics and of the bicharacteristics of the electromagnetic field:

(4.1) $$H := \tfrac{1}{2} \eta^{ik} p_i p_k = 0 \, , \qquad \text{with} \qquad p_i := \frac{\partial z}{\partial y^i} \, ;$$

(4.2) $$\frac{dp_i}{d\sigma} = - \frac{\partial H}{\partial y^i} \, , \qquad \frac{dy^i}{d\sigma} = \frac{\partial H}{\partial p_i} \, ,$$

− where $ds^2 = \eta_{ik} \, dy^i \, dy^k$ gives the Minkowskian interval − play a fundamental role for defining the relativistic notions of time and space. Now, the general relativity includes the special theory as a particular case; therefore for a rational foundation of general theory it is advisable to start from the pseudo-Riemannian generalization of eqs. (4.1) and (4.2). In this way we obtain equations which coincide formally with eqs. (2.1) − (2.1'), (2.2), but which represent the differential equations of the *electromagnetic* wave fronts and rays. Of course, these same equations can be derived from Maxwell equations written for space-time geometry, see e.g. Whittaker (1927) and Fock (1964, p.433).

Remark that also in the light of energy principle Levi-Civita's interpretation of eqs. (2.1) − (2.1'), (2.2) is very reasonable: these equations give indeed the laws of motion of wave fronts and rays which convey the *real* energy of electromagnetic waves, in contrast with the non-localizability of the gravitational *pseudo* energy, which is only a mathematical fiction.

**5.** − *An objection*: The existence of a constant metric along any null ray is inessential to the thesis because also in a Fermi co-ordinate frame there exists in general a nonzero curvature along the ray. Therefore, by arranging a suitable set of observers covering some world region, it should be possible to reveal a gravitational wave. ***The answer***: Of course, Fermi's theorem is not sufficient to exclude the existence of the gravitational waves, but our previous proof (see sect. **2**) rests on the theory of the characteristics of Einstein's field equations, which tells us that – at least for the more important and common kinds of gravitational waves, and for a great portion of any wave at a large distance from the origin – the spatio-temporal wave front is a *developable* hypersurface on which a *constant* metric can be obviously chosen. Consequently, the above mentioned observers are doomed to capture the Nothing.

*A second objection*: The weak-field approximation (Minkowskian metric plus small corrections) seems quite meaningful for various physical situations. An example: an oscillating billiard ball and an observer not very distant from it. For this observer the fact that somewhere far the metric cannot be represented by small corrections to Minkowski tensor seems irrelevant. Localization seems also possible, in principle: in our laboratory we can have the same linear approximation both near the source and the detector. ***The answer***: Contrary to a widespread belief, the *exact* theory does *not* ascribe an absolute meaning to acceleration. Accordingly, nothing is actually emitted by an oscillating billiard ball, by a spinning top, or by whatever device: the emission mechanism suggested by the linear approximation is quite illusive.

*A **third** objection*: Modern gravitational wave production calculations go much beyond the linearized theory. There exist approximation schemes that construct to all orders of perturbation theory the nonlinear radiative gravitational field generated by an arbitrary material. These methods incorporate the no-incoming-radiation condition and show that the asymptotic outgoing field contains outgoing gravitational waves. Such formal perturbative expansions are asymptotic to a sequence of exact solutions of Einstein's theory. (See the perturbative papers by Blanchet and Damour, 1986, Blanchet, 1987 and Damour and Schmidt, 1990).

*The answer*: It is an obvious consequence of the theory of the characteristics that there exist metric fields $g_{ik}(x)$ of a formally undulatory type. But, as we have seen, the same theory proves that the gravitational waves are mere sinuosities generated, in the last analysis, by *wavy reference systems*.

**6.** − The gravitational cylindrical waves of the *rigorous* solution by Einstein and Rosen (1937) yield a good illustration of our thesis. These authors choose a three-dimensional reference frame $x^\lambda$, ($\lambda=1, 2, 3$), such that $x^1=0$ represents the symmetry axis, $x^2$ goes from zero to infinity, and $x^3$ is an angular co-ordinate giving the position of the meridian plane. The cylindrical symmetry requires that only the following components of the fundamental tensor $g_{ik} = g_{ik}(x^1, x^4)$, ($i, k=1, 2, 3, 4$), are different from zero: $g_{11}$, $g_{22}$, $g_{33}$, $g_{44}$, $g_{14}$. With a suitable transformation of $x^1$ and $x^4$, it is also possible to satisfy the conditions $g_{14}=0$ and $g_{11} = -g_{44}$. By solving the field equations $R_{ik} - (1/2)\, g_{ik}\, R = 0$ for empty space, Einstein and Rosen find that there are solutions $g_{11}(x^1, x^4)$, $g_{22}(x^1, x^4)$, $g_{33}(x^1, x^4)$, representing cylindrical waves in a three-dimensional Euclidean *Bildraum* $\mathfrak{E}$ for which $(x^1, x^2, x^3)$, is a usual system of cylindrical co-ordinates and $x^4$ is a usual time parameter. We have in space-time

(6.1) $\qquad ds^2 = g_{\lambda\lambda}(x^1, x^4)\, dx^\lambda\, dx^\lambda - g_{11}(x^1, x^4)\, [dx^4]^2$ .

The corresponding interval in $\mathfrak{E}$ is

(6.2) $\qquad dS^2(T) = g_{\lambda\lambda}(X^1, T)\, dX^\lambda\, dX^\lambda$ ,

where, for conceptual evidence, we have called $X^\lambda$, ($\lambda=1, 2, 3$), the metrical cylindrical co-ordinates, and we have written $T$ in lieu of $x^4$. Obviously, the time independent customary interval $d\Sigma$ in $\mathfrak{E}$ is given by

(6.3) $\qquad d\Sigma^2 = (dX^1)^2 + (dX^2)^2 + (X^1)^2\,(dX^3)^2$ .

The *Bildraum* $\mathfrak{E}$ is flat and therefore we can put

(6.4) $\quad g_{\lambda\lambda}(X^1, T^*) \, dX^\lambda \, dX^\lambda \equiv h_{\lambda\lambda} \, [\xi^1(T^*), \xi^2(T^*), \xi^3(T^*)]$ .

$$. \, d\xi^\lambda(T^*) \, d\xi^\lambda(T^*) = dS^2(T^*),$$

where $T^*$ is any given value of $T$, and the functions

(6.5) $\quad\quad\quad\quad\quad \xi^\lambda(T^*) := f^\lambda(X^1, X^2, X^3, T^*)$

give a suitable system of orthogonal curvilinear co-ordinates in $\mathfrak{E}$. We see – and in a way fully independent of the theory of the characteristics – that the waves $g_{jj}(x^1, x^4)$, ($j$=1, 2, 3, 4), are generated by *a series of co-ordinate changes* in $\mathfrak{E}$: $(X^1, X^1, X^3) \rightarrow [(\xi^1(T), \xi^2(T), \xi^3(T)]$, that depend on the time parameter $T$ and propagate with the velocity of light.

We think that Einstein would not dislike this conclusion: indeed, he had suspected that the gravitational waves are unreal entities. We remember finally, with a famous phrase by Eddington, that an ideal propagation of co-ordinate changes with the *speed of thought* would give origin to an undulating fundamental tensor, whose waves would propagate with the same mental velocity.

ACKNOWLEDGEMENT

I thank cordially Prof. A. Andreatta and Prof. G. Cornalba for a useful discussion on some mathematical points.

APPENDIX

*Nobody has ever found a direct experimental proof of the existence of the gravitational waves*. According to some authors (cf. e.g. Taylor *et al.*, 1979, Taylor and Weisberg, 1982 and Taylor, 1994), an indirect experimental evidence could be given by the time decrease of the orbital period $P_b$ of the famous binary pulsar PSR 1913+16. The measured value of $dP_b/dt$ is $(-2.30\pm0.22)\times10^{-12}$, while the well-known quadrupole formula of the linearized relativity gives a $dP_b/dt$ due to the emission of gravitational radiation equal to $-2.4 \times 10^{-12}$. The agreement is astonishingly good, but rather suspect, as it is clear from our previous considerations; see also Ehlers *et al.*, 1976 and Rosenblum, 1978, who emphasized the unreliability of the linearized theory. An excellent agreement with the observational data has been also obtained by means of computations at third order in $G$ and fifth order in $v/c$. These theoretical computations of $dP_b/dt$ presuppose implicitly that the internal structures of PSR 1913+16 and of its companion are very similar. But if the companion were e.g. a helium star or a white dwarf, its viscous losses could cause a time decrease of $P_b$ of the same order of the one given by the hypothesized emission of gravitational waves.

In conclusion, the belief in the existence of the gravitational radiation is a product of a wishful thinking.

LOINGER A. , ON THE GRAVITATIONAL WAVES## References

Bergmann, P.G.: 1960, *Introduction to the Theory of Relativity,* (Prentice-Hall, Inc., Englewood Cliffs, N.J).

Blanchet, L. and Damour, T.: 1986, *Phil. Trans. R. Soc. London* A **320**, 379.

Blanchet, L.: 1987, *Proc. R. Soc. London* A **409**, 383.

Damour, T. and Schmidt, B.: 1990, *J. Math. Phys.* **31**, 2441.

Ehlers, J., Rosenblum, A., Goldberg, J.N. and Havas, P.: 1976, *Astrophys. J. Lett.* **208**, L77.

Einstein, A. and Rosen, N.: 1937, *J. Franklin Inst.* **223**, 43.

Eisenhart, L.P.: 1927, *Non-Riemannian Geometry* (Am. Math. Soc., New York, N.Y.).

Fermi, E.: 1922, *Rend. Acc. Lincei* **31**/1, 21 and 51.

Fock, V.: 1964, *The Theory of Space, Time and Gravitation*. Second Revised Edition (Pergamon Press, Oxford, etc.).

Levi-Civita, T.: 1925, *Lezioni di calcolo differenziale assoluto* (Stock, Roma).

Levi-Civita, T.: 1930, *Rend. Acc. Lincei* **11** (s.6$^a$), 3 and 113.

Møller, C.: 1972, *The Theory of Relativity*. Second Edition (Clarendon Press, Oxford).

Rosenblum, A.: 1978, *Phys. Rev. Lett.* **41**, 1003.

Straumann, N.: 1984, *General Relativity and Relativistic Astrophysics* (Springer-Verlag, Berlin, etc.).

Taylor, J.H., Fowler, L.A. and McCulloch, P.M.: 1979, *Nature* **277**, 437.

Taylor, J.H. and Weisberg, J.M.: 1982, *Astrophys. J.* **253**, 908.

Taylor, J.H.: 1994, *Rev. Modern Phys.* **66**, 711.

Weyl, H., 1988, *Raum-Zeit-Materie*, Siebente Auflage (Springer-Verlag, Berlin, etc.).

Whittaker, E.T.: 1927, *Proc. Cambridge Phil. Soc.* **24**/1, 32.
*————————————*

page 10 of 10